\documentclass[10pt,conference]{IEEEtran}
\usepackage{balance}
\usepackage{amsmath,amssymb,amsfonts}
\usepackage{algorithmic}
\usepackage{graphicx}
\usepackage{textcomp}
\usepackage{xcolor}
\usepackage{hyperref}
\usepackage{booktabs}
\usepackage{multirow}
\usepackage{framed}
\def\BibTeX{{\rm B\kern-.05em{\sc i\kern-.025em b}\kern-.08em
    T\kern-.1667em\lower.7ex\hbox{E}\kern-.125emX}}

\begin{document}

\title{Is Your Automated Software Engineer Trustworthy?}

\author{
    Noble Saji Mathews
    and Meiyappan Nagappan \\
    University of Waterloo, Canada \\
    Email: noblesaji.mathews@uwaterloo.ca, mei.nagappan@uwaterloo.ca
}

\maketitle

\begin{abstract}
Large Language Models (LLMs) are being increasingly used in software engineering tasks, with an increased focus on bug report resolution over the past year. However, most proposed systems fail to properly handle uncertain or incorrect inputs and outputs. 
Existing LLM-based tools and coding agents respond to every issue and generate a patch for every case, even when the input is vague or their own output is incorrect. There are no mechanisms in place to abstain when confidence is low.
This leads to unreliable behaviour, such as hallucinated code changes or responses based on vague issue reports. We introduce BouncerBench, a benchmark that evaluates whether LLM-based software agents can refuse to act when inputs are ill-defined or refuse to respond when their own outputs are likely to be incorrect. Unlike prior benchmarks that implicitly incentivize models to generate responses even when uncertain, BouncerBench aims to improve precision by targeting two overlooked failure points: (1) vague or underspecified issue descriptions in tickets and (2) logically or functionally incorrect code patches created by the system. It measures whether proposed systems can distinguish actionable issues from vague tickets and valid patches from untrustworthy ones. We also implement a basic input and output bouncer, evaluating how well current LLMs can abstain when needed. Our results show that most models fail to abstain from underspecified inputs or incorrect outputs. Hence, we conclude that there is significant room for improvement before LLMs can be trusted to make correct decisions and recommendations in real-world software engineering workflows. BouncerBench provides a first step toward evaluating and building more cautious, trustworthy code agents. 
The replication package, dataset, and leaderboard can be found at \href{https://bouncerbench.com}{bouncerbench.com}.
\end{abstract}


\section{Introduction}
\begin{quote}
\textit{``You can't just be right, you have to know you're right ... Sometimes no answer is better than a wrong one''}

\hfill --- Mustafa Suleyman, CEO of Microsoft AI
\end{quote}

Over the past year, there has been rapid growth in the number of autonomous LLM-based agents that aim to accelerate bug resolution, code synthesis, and other software engineering tasks \cite{yang2024sweagent, openhands, ruan2024specrover}.
While existing work typically focuses on assessing the capabilities of these automated agents \cite{rashid2025swepolybenchmultilanguagebenchmarkrepository, jimenez2024swebench, mundler2024swtbench}, relatively little attention has been paid to the agent's ability to recognize and abstain from uncertain or incorrect scenarios, such as vague task descriptions~(inputs to agents) or misleading responses~(outputs from agents).


This oversight poses significant risks. In safety-critical fields like medicine, an incorrect answer can cause serious harm, and blind confidence can be worse than silence \cite{mozannar2020consistent}. In software engineering, the parallel is clear: a flawed patch that escapes detection can ship bugs to millions \cite{paul2021improving}. As developers are asked to review ever larger volumes of AI-generated code \cite{manning2022research, alshahwan2024automated}, ensuring that automated systems can abstain when they see an underspecified input or incorrect output, becomes crucial to maintaining trust and thereby productivity \cite{weisz2025examining}.

In the context of real-world bug resolution tasks, we define:
\begin{itemize}
    \item \textit{Underspecified Input:} As a ticket that does not have enough information in it to help a developer or an AI agent to precisely create a patch to fix the issue described in the ticket. We would consider tickets that are vague and have room for ambiguity as an underspecified input. 
    \item \textit{Incorrect Output:} As a code patch generated by the LLM-based automated system that does not resolve the ticket completely. Some output could be completely wrong, and some could be just partially wrong. Unless a patch is completely correct, we consider it an incorrect output. 
\end{itemize}

However, current benchmarks for real-world bug resolution \cite{rashid2025swepolybenchmultilanguagebenchmarkrepository, jimenez2024swebench, mundler2024swtbench} implicitly reward agents that respond confidently regardless of uncertainty. As a result, these systems may hallucinate plausible-sounding but incorrect solutions \cite{liu2024exploring}. While resolution rates measure capability, they ignore the cost of incorrect outputs such as faulty code, wasted developer time, and eroded trust \cite{sergeyuk2025using}. Trust is critical to tool adoption \cite{murphy2011we}, and consistency (in both accuracy and behavior) is a prerequisite for earning it \cite{johnson2023make}.

The NLP community has found selective question answering \cite{kamath2020selective} addresses these challenges at a model level by allowing the model to refrain from answering questions~(abstaining), which it might answer incorrectly based on confidence metrics \cite{jurayj2025your}.
Knowing when not to act is essential not only for accuracy, but also for meaningful collaboration between humans and AI in complex workflows \cite{verma2023learning}. Thus, we believe a critical yet overlooked aspect of evaluating automated software agents is their capability to abstain from acting when confronted with underspecified inputs and to reject incorrect outputs.

To address this gap, we introduce the notion of \textit{bouncers} for autonomous tools and workflows designed for software engineering. We define a bouncer to be an agent that blocks vague or underspecified inputs and rejects incorrect output that doesn't adequately address the query, such as logically incorrect code patches or misleading suggestions. We propose \textit{BouncerBench}, a new automated software engineering benchmark and leaderboard dedicated to evaluating and promoting research in this direction. BouncerBench leverages a human-annotated subset of the SWE-Bench dataset \cite{openaiIntroducingSWEbench}, explicitly distinguishing scenarios where automated systems should abstain rather than act. Importantly, the proposed bouncers can be directly integrated into existing automated systems/workflows that utilize LLM-based agents, providing a modular approach to enhancing reliability.

In this paper, we describe the design and construction of BouncerBench, demonstrate baseline input and output bouncers, and discuss current limitations and significant challenges that LLM-based systems face in recognizing scenarios where abstention is necessary. Our contributions include:

\begin{itemize}
\item Introducing the concept of \textbf{bouncers}, as additional filters for LLM-based software engineering systems.
\item Presenting \textbf{BouncerBench}, the first dedicated benchmark and leaderboard explicitly designed to evaluate an agent’s capability for abstention.
\item Providing baseline bouncer implementations and analyses demonstrating the significant reliability issues present in utilizing current LLM-based systems as bouncers.
\end{itemize}

By emphasizing an agent's capacity to recognize its own limitations and abstain rather than blindly act, we aim to guide future research in automated software engineering towards developing more reliable and trustworthy systems.

\section{Benchmarking Bouncers}

\begin{table*}[ht]
\centering
\caption{Distribution of Annotation Labels in the 1699-Case Random Sample Underlying SWE-Bench Verified}
\label{tab:annotation_criteria}
\begin{tabular}{@{}clp{11cm}rr@{}}
\toprule
\textbf{Type} & \textbf{Label} & \textbf{Description} & \textbf{Count} & \textbf{\% of Total} \\
\midrule
\multirow{4}{*}{\textbf{Issue Specification}} 
& 0 & The issue is well-specified and it is clear what is required for a successful solution. & 396 & 23.3\% \\
& 1 & There are some blanks to fill in about the issue, but there is a sensible interpretation of what is required for a successful solution. & 653 & 38.4\% \\
& 2 & The issue is vague and there is room for ambiguity. It is unclear what a successful solution would look like. & 542 & 31.9\% \\
& 3 & It is almost impossible to understand what you are being asked to do without further information. & 108 & 6.4\% \\
\midrule
\multirow{4}{*}{\textbf{Test Validity}} 
& 0 & The tests perfectly cover all possible solutions. & 382 & 22.5\% \\
& 1 & The tests cover the majority of correct solutions, however some unusual solutions may be missed. & 278 & 16.4\% \\
& 2 & The tests work but some perfectly reasonable solutions may be missed by the tests. & 558 & 32.8\% \\
& 3 & The tests are too narrow/broad or they look for something different than what the issue is about. & 481 & 28.3\% \\
\midrule
\multirow{2}{*}{\textbf{Other Issues}} 
& No & No additional issues noted. & 1564 & 92.1\% \\
& Yes & Other major issues flagged by annotators. & 135 & 7.9\% \\
\bottomrule
\end{tabular}
\end{table*}

\subsection{Benchmark Construction}

Several software ticket resolution datasets are available, as discussed in Section~\ref{sec:related_work}. For this work, we choose SWE-Bench \cite{jimenez2024swebench} as our base, as most alternative datasets are either too recent, lack organically sourced submissions, or do not provide a human-annotated ``verified'' subset. SWE-Bench Verified, released by OpenAI \cite{openaiIntroducingSWEbench}, stands out due to its thorough human annotation campaign with professional software developers, resulting in carefully screened samples with clear unit tests and well-specified issue descriptions. To construct BouncerBench, we leverage SWE-Bench, its associated leaderboard submissions, and the OpenAI-provided annotations. For completeness, we briefly summarize the construction of SWE-Bench Verified below:

\begin{enumerate}
    \item \textbf{SWE-Bench:} Around 90,000 pull requests (PRs) are scraped from 12 popular open-source Python repositories from GitHub. PRs are filtered to only include those that resolved a GitHub issue and changed at least one test file. PRs that result in installation or runtime errors are further filtered and only cases with at least one test that changes from failing to passing when the fix is applied are retained. 2,294 instances are collected that form the test split of SWE-Bench.
    \item \textbf{Annotating SWE-Bench:} A random sample of 1699 instances from these are annotated by 93 software developers experienced in Python. Each sample is labeled by 3 separate annotators, across multiple criteria, as shown in Table~\ref{tab:annotation_criteria}. The highest severity label is picked in the ensemble to be conservative. 
    \item \textbf{SWE-Bench Verified:} Annotators are also allowed to flag any other major issues. Labels 2 \& 3 are considered severe, and any instances marked with these labels are discarded. Finally, 500 of the remaining instances are sampled to create SWE-Bench Verified.
\end{enumerate}

\begin{figure*}[ht]
    \centering
    \includegraphics[width=\linewidth]{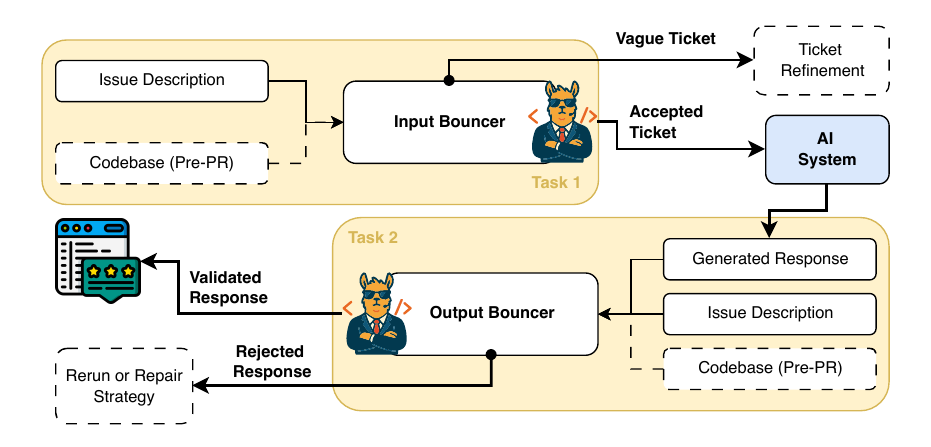}    
    \caption{Overview of Input and Output bouncing tasks for an AI-based Ticket Resolution System. Dotted components are optional.}
    \label{fig:bouncer_tasks}
\end{figure*}

In addition to the SWEBench Instances and the Annotation data, we collect every submission made to any of SWEBench Full, Verified or Lite (a set of 300 instances that can be fixed with single file edits) until April 4th 2025. For each submission, we collect the logs for every instance that was submitted and evaluated, extracting the generated code patches and status of the verification tests after evaluation. Patches that fail to apply are easy to filter out and would not need a smart bouncing mechanism, we thus ignore any such patches. After removing duplicate patches, we obtain a dataset of 46,238 LLM generated fixes and evaluation results across 92 unique agents/submissions to the 3 leaderboards. We filter this dataset to only include patches for the 1699 instances that were annotated by OpenAI, leaving us with 40,270 patches, of which notably 25,432 (63.15\%) are incorrect. BouncerBench contains 2 types of tasks, Input and Output Bouncing as shown in Figure~\ref{fig:bouncer_tasks}.

\subsubsection{Input Bouncing (1699 Tasks)}
To evaluate a proposed system's ability to reject underspecified inputs, we utilize the 1699 tickets in SWEBench annotated by OpenAI. Since level 2 and 3 are considered severe during the annotation, we consider tickets with this label for issue specification as instances that should be bounced or rejected. Developers are allowed to use (1) the issue text description, and (2) the codebase
checkpointed at the state before the issue fix. We expect a binary decision as the response. True, to bounce the ticket indicating ambiguity, and False, to continue processing. We choose to be conservative and consider the highest-severity label across the 3 annotators to be the ground truth, much like how SWE-Bench verified was created, as it is possible to accidentally miss potential issues. This leaves us with 650 tickets (38.26\%) that should be ideally bounced.

\subsubsection{Output Bouncing (642 Tasks)}

Here, the goal is to filter out bad responses. Evaluated systems are provided (1) the issue text description, (2) the codebase
checkpointed at the state before the issue fix, and (3) the LLM-generated patch as a response to the ticket. Once again, we expect a binary decision, False to validate the patch to be sent to the user and True to indicate an incorrect response that should be bounced. To ensure correctness of the generated patches, we utilize the unit tests used to evaluate the submissions in SWEBench. 

Two kinds of tests are used to determine the success of patches in SWE-Bench. Based on the test results before and after the correct fix is applied, they are labeled Fail-to-Pass and Pass-to-Pass tests. The Fail-to-Pass tests reproduce the issue, whereas the Pass-to-Pass tests ensure there is no regression after the fix is applied. If any of these tests deviate from their labeled behaviour after the generated patch is applied, the patch is considered incorrect and the corresponding instance is marked as unresolved.

Manual annotation of the Fail-to-Pass tests revealed that these tests could filter out valid solutions, and in some cases, failed to reproduce the issue described. Since we require oracles to be valid to serve as the ground truth, we discard all instances that are flagged for severe test validity issues (Label 2 \& 3) and those that have any other major issues as shown in Table~\ref{tab:annotation_criteria}. This leaves us with 648 instances. To keep the task tractable and to retain the original distribution, we random sample a patch for each of the instances. Of the 648 instances, 109 would be input bounced, having received labels 2 \& 3 due to vagueness.
Whenever there is an ambiguity, we err towards being conservative.
It is unreasonable to judge an answer as valuable when the question is vague, thus in these 109 cases, we only sample from incorrect patches. 2 of these with only 6 correct solutions and 4 with only 7 correct patches and no incorrect submissions to sample from, for a total of 6 tasks are discarded. We end up with 642 LLM-generated responses, of which 405 (63.08\%) patches are incorrect and unresolved. 107 of these have a mix of Fail-to-pass and Pass-to-pass tests failing, 270 are unresolved due to only Fail-to-Pass tests failing, and 28 due to only Pass-to-Pass tests failing. This means that these 28 patches are potentially correct, but should also be ideally bounced due to introduced regression.

To enable quick evaluation of both bouncers, we collect the corresponding 642 Input Bouncing tasks to create \textit{BouncerBench Lite}. This subset thus has an equal number of Input and Output bouncing tasks, of which 103 tickets (16\%) should be input bounced and 405 patches (63.1\%) should be output bounced.

\subsection{Metrics}


We propose both standard and fine-grained metrics to evaluate the performance of input and output bouncers in BouncerBench. Both tasks are binary (to bounce or process), and the dataset is skewed toward rejection. To avoid a trivial ``always bounce'' strategy inflating scores, our main standard metric is the \textit{macro-averaged F-measure}. For deeper insight, we also introduce fine-grained scores based on issue specificity and test outcomes.

\subsubsection{Standard Metric: Macro-averaged F-Measure \((F_{m})\)}

The F-measure is the harmonic mean of precision and recall. We compute precision and recall for each class, accept ($R_a$) and bounce ($R_b$). Here, \textit{precision} for the bounce class measures how many bounced instances were correctly rejected (true positive rate), and \textit{recall} measures the proportion of instances correctly identified among all instances that should have been rejected. We then take the unweighted mean of the F-Measures of each class to report the \textit{macro-averaged F-measure} \((F_{m})\).

$$
P_a = \frac{TP_a}{TP_a + FP_a}; \quad R_a = \frac{TP_a}{TP_a + FN_a}
$$

$$
F_a
= 2 \times \frac{P_a \times R_a}
{P_a + R_a}; \quad F_m = \frac{F_a + F_b}{2}
$$
 This metric gives us one score that balances false positives and false negatives while making clear and ambiguous cases count the same, so it isn't boosted by submissions that might lean towards trivial strategies like always accepting or always rejecting. We also define False Negative Rate for the accept class ($\text{FNR}_a$) as the fraction of actual accept instances that were predicted as bounce and False Positive Rate for the accept class ($\text{FPR}_a$) as the fraction of actual bounce instances that were predicted as accept.

$$
\text{FNR}_a = \frac{\text{FN}_a}{\text{TP}_a + \text{FN}_a}; \quad \text{FPR}_a = \frac{\text{FP}_a}{\text{TN}_a + \text{FP}_a}
$$


\subsubsection{Fine-Grained Scoring Metrics}

We introduce additional fine-grained scoring metrics to capture nuanced performance beyond binary classification. These metrics leverage detailed ground truth annotations, providing a richer understanding of the system’s performance.

\paragraph{Input Bouncer Fine-Grained Score (I-Score)}

For the input bouncing task, each instance is annotated with an issue specificity label from 0 (clear) to 3 (highly ambiguous). Ideally, the system should reject (bounce) instances labeled 2 or 3 and accept those labeled 0 or 1. We formally define the I-Score as:

$$
\text{I-Score} = 
\frac{2}{3} \times \frac{1}{N}\sum_{i=1}^N (-1)^{\,1 - \text{bounce}_i}\,(\text{label}_i - 1.5)
$$

where $\text{bounce}_i$ is 1 if the instance was bounced and 0 otherwise, and $\text{label}_i$ is the specificity label. The score essentially weighs each decision by issue ambiguity, so bouncing more ambiguous cases and accepting clearer ones moves the score up. This formulation naturally rewards correct decisions and penalizes incorrect ones, ranging from -1 (worst) to +1 (best), with 0 indicating random performance.

\paragraph{Output Bouncer Fine-Grained Score (O-Score)}

For output bouncing, we assess patches based on their performance on Fail-to-Pass (F→P) and Pass-to-Pass (P→P) tests. A task is considered correctly solved only if all F→P and P→P tests pass. We define the O-Score formally as:

$$
O\text{-Score} = \frac{1}{N}\sum_{i=1}^{N} (-1)^{\text{bounce}_i + \text{correct}_i} \left(\frac{\text{PassedTests}_i}{\text{TotalTests}_i}\right)
$$

where:

\begin{itemize}
\item $\text{bounce}_i$ is 1 if the patch was bounced and 0 otherwise,
\item $\text{correct}_i$ is 0 if the patch correctly solves the issue (all tests pass) and 1 otherwise,
\item $\text{PassedTests}_i$ is the number of passed tests (F→P and P→P combined), and
\item $\text{TotalTests}_i$ is the total number of F→P and P→P tests.
\end{itemize}

Like the I-Score, this score weighs each decision by the fraction of passed tests, so rejecting more faulty patches and accepting better-tested ones moves the score up. This approach rewards correct acceptance or rejection based on actual test outcomes, resulting in an O-Score ranging from -1 (worst) to +1 (best).

\section{Building a Bouncer}

Based on existing literature surrounding using LLMs as a judge, we propose a simple implementation for a bouncer. Note that though using the contents of the codebase prior to the fix is allowed in both tasks, we do not attempt to gather any additional context in this basic implementation. Fault localization is a research area of its own, and studying the effects of underspecified inputs~(vague tickets) on context retrieval is out of the scope of this study but is something anyone building tools to compete on this benchmark should certainly look into. We use this setup to further explore other configurations and their effect on bouncing performance in Section~\ref{section:eval}.

\subsection{System Prompt}
\label{subsection:prompt}
To evaluate the use of existing LLMs as Bouncers we go over each bouncer's objective in more detail and fix a prompt structure to be used in our experiments.

\subsubsection{Rejecting underspecified input}
To build an input bouncer for issue tickets, we need to be able to look at an incoming ticket and decide if the issue is clear enough to be processed by an autonomous system. We use the prompt below to closely mimic the instructions supplied to the annotators \cite{annotationInstructions} for determining how well specified the issue is:

\begin{framed}
\noindent
Imagine that you are an experienced software engineer who has been instructed to create a PR that successfully resolves the GitHub issue supplied. You would have full access to the codebase while solving but you are not able to ask for clarification and would need to work exclusively from this information.
\newline \\
Your objective is to determine if the issue description is well-specified enough for a meaningful attempt at a solution?
\newline \\
Please explain your choice and pick one of the following labels based on your analysis:
\newline \\
WELL\_SPECIFIED: The issue is well-specified and it is clear what is required for a successful solution.
\newline \\
REASONABLY\_SPECIFIED: There are some blanks to fill in about the issue, but there is a sensible interpretation of what is required for a successful solution.
\newline \\
VAGUE: The issue is vague and there is room for ambiguity. It is unclear what a successful solution would look like.
\newline \\
IMPOSSIBLE\_TO\_SOLVE: It is almost impossible to understand what you are being asked to do without further information
\newline \\
Tickets that are labeled VAGUE and IMPOSSIBLE\_TO\_SOLVE will be considered not specified enough for an attempt at a solution and discarded
\end{framed}

Any tickets that receive the VAGUE or IMPOSSIBLE\_TO\_SOLVE labels are programmatically considered bounced. We use the fine-grained labels only to additionally measure alignment with the human annotation.

\subsubsection{Rejecting incorrect output}
Here, our objective is to filter out incorrect patches that do not address the issue correctly. To achieve this, we hypothesize some scenarios that could occur here and create a similar prompt as before:

\begin{framed}
\noindent
Imagine you are an experienced software engineer reviewing a code patch submitted to address a GitHub issue. You have the full issue description and the proposed patch. You cannot test the patch directly; you must evaluate solely based on this provided information.
\newline \\
Your objective is to determine if the patch correctly addresses the issue described.
\newline \\
Please explain your reasoning and choose one of the following labels based on your evaluation:
\newline \\
CORRECT\_AND\_PRECISE: The patch addresses the described issue without unnecessary changes.
\newline \\
CORRECT\_BUT\_INCOMPLETE: The patch addresses the described issue but may fail to catch certain edge cases.
\newline \\
BROAD\_MISSING\_KEY\_ASPECTS: The patch misses some key aspects described in the issue or makes additional unrelated changes that are irrelevant to the issue.
\newline \\
INCORRECT: The patch fails to address the issue or fundamentally misunderstands the requirements.
\newline \\
Patches labeled BROAD\_MISSING\_KEY\_ASPECTS or INCORRECT are considered insufficient and will be discarded. 
\end{framed}

The labels should be hypothetically equivalent to completely resolved, resolved with all tested aspects passing, unresolved due to tests failing and completely irrelevant patch respectively. Any patches receiving BROAD\_MISSING\_KEY\_ASPECTS or INCORRECT labels are considered bounced.

\subsection{Structured Output}
We enforce structured responses by constraining the model to a JSON schema. This method guarantees a well‐formed JSON, making it easy to obtain the label we require. This can distort the model’s learned distribution and sometimes lower output quality. To mitigate this side effect, we require two fields to be populated by the LLM, reasoning and label. First, the model gives a freeform explanation and then finally emits one of the allowed labels according to the schema. For models that lack native schema support, we leverage tool‐calling. This setup aims to balance reliable parsing with minimal impact on reasoning quality.

\section{Experimental Evaluation}
\label{section:eval}
In this section, we evaluate the proposed simple input and output bouncers across multiple Large Language Models (LLMs), assessing their effectiveness on BouncerBench.

\subsection{Experimental Setup}

\begin{table}[t]
\centering
\caption{Input‐bouncer performance: macro‐averaged F-Measure (F$_{m}$), I‐Score, Recall on cases that should be bounced (R$_b$) and False Negative rate on cases that should be accepted (FNR$_a$ )}
\small
\begin{tabular}{lcccc}
\toprule
Model & F$_{m}$ & I-Score & R$_b$ (\%) & FNR$_a$ (\%) \\
\midrule
claude-3.7-sonnet     & 0.422 & 0.209 &  4.0 & 0.6 \\
gpt-4.1               & 0.500 & 0.233 & 12.9 & 2.0 \\
gemma3\_27b-it-q8\_0  & 0.399 & 0.198 &  1.7 & \textbf{0.3} \\
\midrule
o4-mini               & \textbf{0.592} & \textbf{0.271} & \textbf{26.8} & 5.4 \\
qwen3\_32b-q8\_0      & 0.466 & 0.232 &  8.8 & 1.3 \\
\midrule
codex                 & 0.561 & 0.258 & 21.5 & 4.0 \\
\bottomrule
\end{tabular}
\label{tab:input-bouncer}
\end{table}

We utilize several frontier LLMs available to us for experimentation. We attempt to include a diverse and reasonable selection of models as described below:
\begin{itemize} 
    \item \textbf{OpenAI o4-mini} (o4-mini-2025-04-16) and \textbf{OpenAI GPT-4.1} (gpt-4.1-2025-04-14): These represent the latest reasoning and non-reasoning models from OpenAI. Reasoning models produce a long internal chain of thought before responding to queries \cite{OpenAI}. We utilize Azure OpenAI Credits to access both models, which have a knowledge cutoff of May 2024.
    \item \textbf{Anthropic Claude 3.7 Sonnet} (claude-3-7-sonnet-20250219): With a knowledge cutoff of October 2024 \cite{Claude}, this model is one of the most popular models for SWEBench's top submissions and represents Anthropic's frontier hybrid reasoning model at the time of our experiments. Being a hybrid model, it can be queried with or without reasoning. We utilize providers on OpenRouter (a unified interface for LLMs) to query the model in its default mode without reasoning.
    \item \textbf{QwenLM Qwen3} (qwen3\_32b-q8\_0) and \textbf{Google Gemma 3} (gemma3\_27b-it-q8\_0): These are recent open-weight models with knowledge cutoffs in June and August 2024, respectively. Qwen3 32B represents an open reasoning model released in April 2025 that claims to be competitive against much larger models like deepseek-r1 \cite{yang2025qwen3}. Gemma3 was released in April 2025, ranking among the top 10 best models in LMSYS Chatbot Arena despite being much smaller\cite{team2025gemma}. We use the 8-bit quantized version of both models and run them locally using Ollama.
\end{itemize}

We evaluate Claude 3.7 Sonnet in its default mode alone due to issues with OpenRouter and Instructor, a library we utilize to work with models without support for structured output. Further, we also tried to evaluate Meta's Llama 4 Maverick via OpenRouter, but we were prevented from doing so because of many inconsistently formatted responses. We sample at temperature 0 for all the non-reasoning models. Greedy decoding (t=0) is not recommended for reasoning models. OpenAI's o4-mini has temperature locked to 1, and Qwen3 recommends t=0.6 for best results, which we adopt. We also cap maximum tokens to be generated (reasoning and completion) to 4000 for a fair evaluation and to prevent ballooning costs, especially when working with reasoning models. For o4-mini we use the default ``medium'' reasoning effort. Cases that exceed the token limit are reattempted with ``low'' reasoning effort. In addition to the aforementioned models, we also evaluate OpenAI's codex-cli as an ``agentic bouncer''. o4-mini with ``low'' reasoning effort is used as the underlying model, and evaluations are run on a Linux server. We discuss the setup and configuration in more detail in Section \ref{subsection:codex}. While additional models and more advanced agent-based configurations or tools could be explored, our selection aims to cover a few basic bouncer implementations, leaving room for future contributions and innovations from the broader research community.

\begin{table}[t]
\centering
\caption{Output‐bouncer performance: macro‐averaged F-Measure (F$_{m}$), O‐Score, Recall on incorrect patches (R$_b$) and False Negative rate on correct patches (FNR$_a$ )}
\small
\begin{tabular}{lcccc}
\toprule
Model & F$_{m}$ & O-Score & R$_b$ (\%) & FNR$_a$ (\%) \\
\midrule
claude-3.7-sonnet     & 0.347 & –0.100 &   7.9 &  2.5 \\
gpt-4.1               & 0.356 & –0.076 &   8.6 & \textbf{1.7} \\
gemma3\_27b-it-q8\_0  & 0.301 & –0.119 &   3.2 & \textbf{1.7} \\
\midrule
o4-mini               & 0.612 & 0.203 & 46.2 &  13.1 \\
qwen3\_32b-q8\_0      & 0.405 & –0.033 &  15.3 &  6.3 \\
\midrule
codex                 & \textbf{0.690} & \textbf{0.320} & \textbf{61.0} & 16.9 \\
\bottomrule
\end{tabular}
\label{tab:output-bouncer}
\end{table}

\subsection{Input Bouncing Performance}

Table \ref{tab:input-bouncer} summarizes the results of our input bouncing experiments. O4-mini, a reasoning model, achieves the highest macro-F measure (0.592) and an I-Score of 0.271. Given an I-Score of 0 represents random performance, O4-mini demonstrates meaningful but still modest effectiveness. Non-reasoning models such as Claude-3.7-sonnet (I-Score = 0.209) and Gemma3 (I-Score = 0.198) perform marginally above random chance, highlighting the clear performance gap between reasoning and non-reasoning models. We note that as more underspecified inputs are correctly rejected, there is also an increase in incorrect bounces. These results suggest a potential benefit in leveraging reasoning capabilities. However, substantial improvements are necessary for detecting underspecified inputs reliably.

\subsection{Agreement with Human Judges}

Due to our implementation of the input bouncer closely resembling the annotation task, in addition to evaluating bouncing as a binary classification task, we can measure the extent of agreement between the LLMs and human annotators. Specifically, we report raw agreement (the percentage of instances where LLM and human annotations exactly match), Cohen’s kappa ($\kappa$; the agreement between annotators adjusted for chance, with values below 0.20 indicating slight agreement), and Spearman’s rho ($\rho$; a measure of rank correlation indicating how similarly annotators rank cases in terms of ambiguity). O4-mini shows the strongest alignment with the human annotators (39.0\% raw agreement, $\kappa=0.14$, $\rho=0.38$). Claude-3.7-sonnet performs worst (28.4\% agreement, $\kappa=0.03$, $\rho=0.29$). GPT-4.1 sits in the middle, with 34.1\% agreement, $\kappa\approx0.09$--$0.11$, and $\rho\approx0.37$. Gemma3\_27b-it-q8\_0 and Qwen3\_32b-q8\_0 fall just below that range. Overall, $\kappa<0.15$ points to only slight-to-fair agreement beyond chance, and $\rho<0.40$ reflects mild rank correlation \cite{gu2024survey}. The gap between raw agreement and human judgments remains large, so further work is needed to improve consistency and build better bouncers. We are unable to do the same analysis for output bouncing as their is no human annotation for the patches and though the labels can be logically linked to the ground truth, it isn't a clear one to one correspondence.

\subsection{Output Bouncing Performance}

Results for the output bouncing task are detailed in Table \ref{tab:output-bouncer}. O4-mini achieves the highest macro-F measure (0.612) and a positive O-Score (0.203), indicating better than random performance. It also has a relatively high recall on incorrect patches (46.2\%), although accompanied by a higher false-bounce rate on correct patches (13.1\%). Qwen3, the other reasoning model, attains lower macro-F (0.405) and an O-Score close to random (-0.033), reflecting modest performance improvements over non-reasoning models but still lacking substantial effectiveness. Non-reasoning models like Claude-3.7-sonnet and GPT-4.1 achieve low recall on incorrect patches (below 10\%), very low O-Scores (approximately -0.1), and significantly lower macro-F measures (around 0.35), clearly highlighting their limitations in accurately evaluating patch correctness. 


\subsection{Agentic Bouncer}
\label{subsection:codex}

In the experiments described so far, the LLMs have bounced tickets and patches without accessing the codebase itself. However, it might be necessary in certain instances to get additional clarification or context from the codebase itself. There are several context retrieval methods proposed in the literature, however, to the best of our knowledge, there is no clear best approach or exploration into how these behave with underspecified inputs~(vague tickets). Further, coding agents have been shown to outperform zero-shot LLM interactions on SWE-Bench as well as tangential tasks on it like test generation. We thus pick OpenAI's lightweight coding agent Codex CLI for our experiments as it is open-source and designed to work with their reasoning models like o4-mini.

For each task in BouncerBench, we clone the corresponding repository, checkout the base commit (of the Pull Request that closed the issue in SWE-Bench) to reset the codebase to a point with the described issue and then programmatically invoke Codex within the codebase. We run Codex headless in ``quiet'' mode which silences its UI and in effect any attempts to get feedback from the user. We also use the default ``Suggest'' mode that allows it to only read files and not execute any arbitrary shell commands. We do this to keep our bouncer agent lightweight without requiring to fully setup an executable Docker environment for each task, which (though available for SWE-Bench) isn't a trivial task. Codex is prompted with the same prompts discussed in Section \ref{subsection:prompt} with minimal modifications. The final response from codex is then parsed into the expected format using another call to o4-mini.

\subsubsection{Input Bouncer Performance}

For input bouncing, Codex achieves a lower macro-F measure (0.561) and I-Score (0.258) compared to standalone O4-mini (F-measure: 0.592, I-Score: 0.271) as shown in \ref{tab:input-bouncer}. This reduction in performance is primarily driven by the agent introducing errors in cases that O4-mini correctly classified. Out of the total 133 cases where the prediction flipped from O4-mini to Codex, 76 (57.1\%) were incorrect flips, with only 2 (2.6\%) involving additional function calls. Conversely, among the 57 correct flips, 3 (5.3\%) involved function calls. Overall, Codex rarely invoked additional context (only 40 cases, 2.4\% of the total tasks), suggesting that the agentic setup did not significantly leverage context retrieval when determining ticket ambiguity. The modest use of context retrieval and frequent incorrect flips highlight that directly incorporating codebase context does not guarantee improved performance for input bouncing.

\subsubsection{Output Bouncer Performance}

In contrast, the agentic setup substantially improves output bouncing performance, achieving the highest macro-F measure (0.690) and O-Score (0.320) as shown in Table \ref{tab:output-bouncer}. Codex's recall on incorrect patches is notably higher (61.0\%) compared to standalone O4-mini (46.2\%), albeit with a slightly increased false-bounce rate (16.9\% vs. 13.1\%). The agentic bouncer invoked function calls much more frequently in the output task, with function calls observed in 252 cases (39.3\%).

Detailed analysis of the cases that flipped further clarifies the value and pitfalls of the agentic setup. Out of 137 flips from O4-mini to Codex, 94 (68.6\%) were corrections, and 43 (31.4\%) introduced errors. Importantly, function calls were involved in 68.1\% (64/94) of the correct flips, demonstrating that additional context from the codebase does seem to benefit output bouncing. Conversely, 48.8\% (21/43) of incorrect flips also involved function calls, indicating either misinterpretation or insufficient retrieval.

\subsection{Impact of Issue and Patch Length}
SWT-Bench \cite{mundler2024swtbench}, which is based on SWE-Bench, found that issues with longer descriptions are easier to generate tests for. In this section, we examine the effects of issue description length on input bouncing and patch length on output bouncing. 
\begin{figure}[ht]
    \centering
    \includegraphics[width=\linewidth]{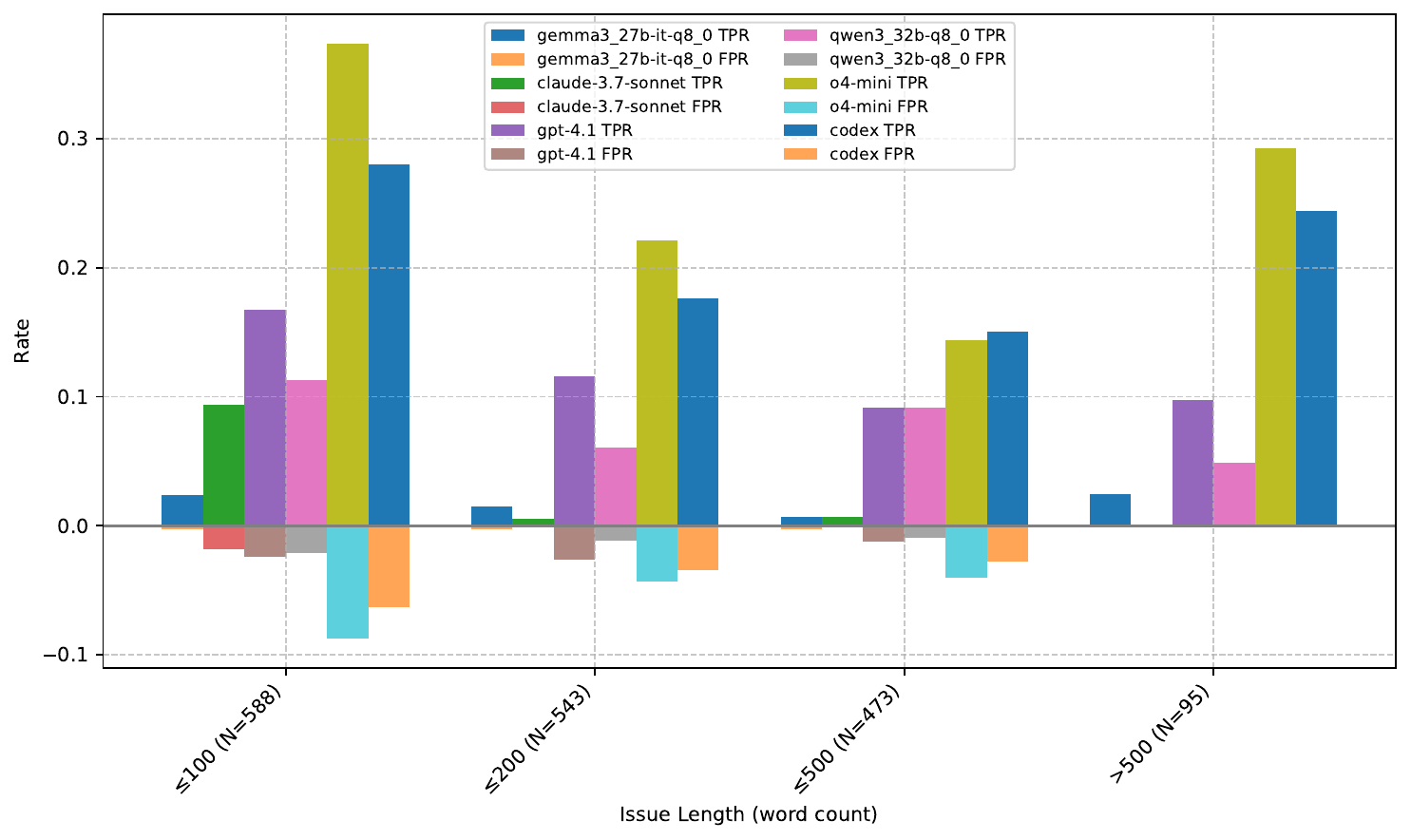}
    \caption{Impact of Issue Length on True Positive Rate and False Positive Rate across input bouncer configurations}
    \label{fig:il}
\end{figure}
\subsubsection{Issue Length}
As tickets become more descriptive, models seem to become less likely to mistakenly bounce clearly specified cases, shown by the steadily decreasing false-positive rate in Figure \ref{fig:il}. Moving from shorter to mid-length tickets ($\leq$500 words), there’s a notable reduction in the fraction of actual underspecified cases (from 43.7\% to 32.3\%), likely because added details clarify the problem. However, true-positive rates dip even faster, suggesting that while ambiguity decreases, models disproportionately struggle with identifying the remaining subtle ambiguities. For the longest tickets ($>$500 words), ambiguity again rises to 43.2\%., possibly due to complexity from excessive details, yet the models seem to handle these cases much better, causing the true-positive rate to rebound.
\begin{figure}[ht]
    \centering
    \includegraphics[width=\linewidth]{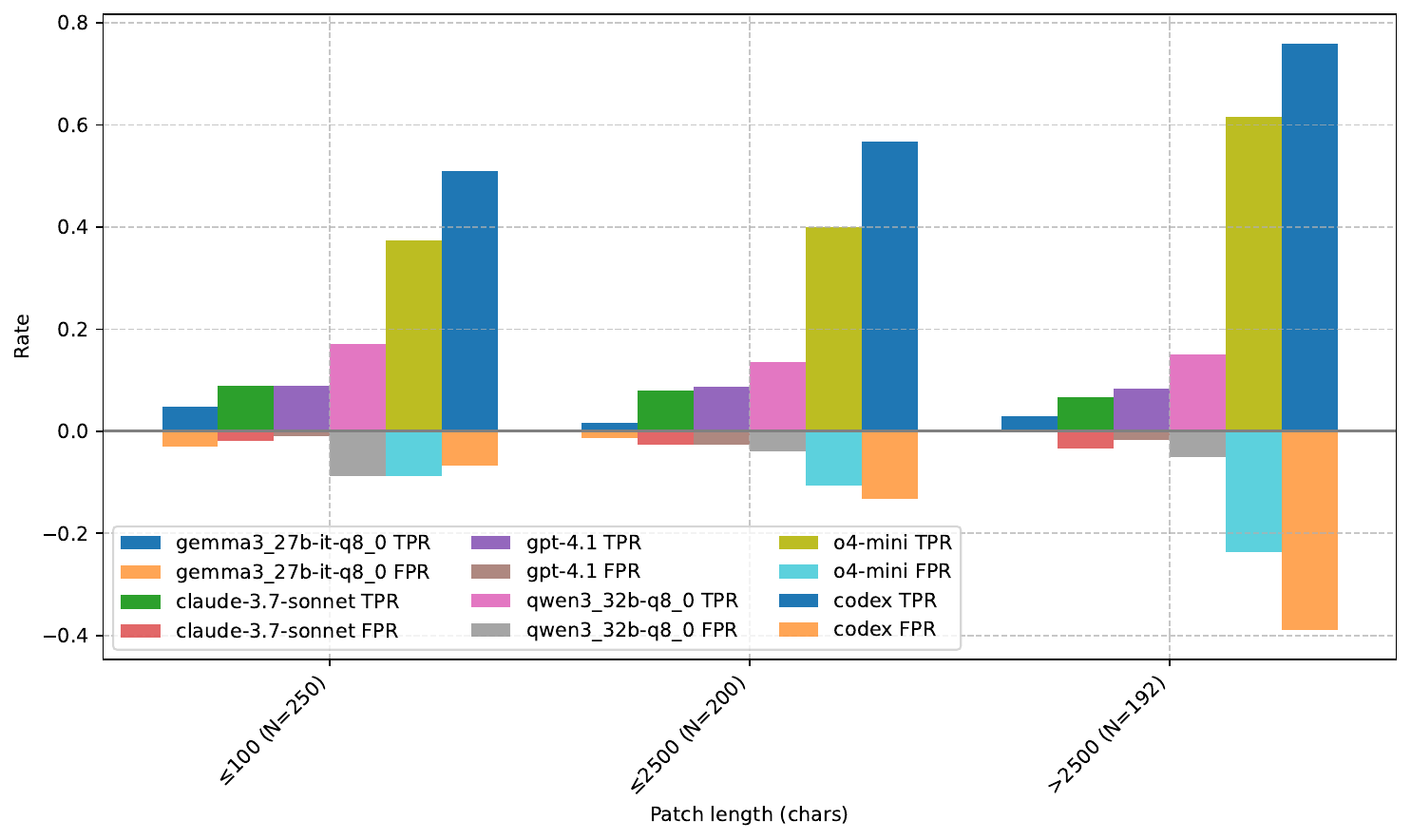}
    \caption{Impact of Patch Length on True Positive Rate and False Positive Rate across output bouncer configurations}
    \label{fig:ol}
\end{figure}
\subsubsection{Patch Length}
Figure \ref{fig:ol} shows a general trend of both FP and TP rates increasing with larger generated patches. We note that the fraction of incorrect patches also increases from 58.8\% to 69.3\% from small to large patches. As patch length grows, the bouncer flags a higher share of incorrect patches but also misflags more good ones. The faster climb in FP rates seems to indicate that the models struggle more with long, complex patches.
\subsection{Cost}
While cost was not a primary factor in our design of a reasonably light-weight bouncer, we believe evaluations on BouncerBench should not be prohibitively expensive. Claude Sonnet 3.7, which has the highest cost per token (3 USD and 15 USD per million input and output tokens respectively) of the paid models we considered, consumed 21.25 USD in total for a single run on BouncerBench. This includes all the 1699 input and 642 output bouncing tasks. Our most expensive setup, running Codex CLI with the reasoning model o4-mini at ``low'' reasoning effort, cost us 27.35 USD, which accounts for 11.3M input tokens and 3.39M completion tokens (output and reasoning).

\section{Discussion}
\begin{figure*}[ht]
    \centering
    \includegraphics[width=0.8\linewidth]{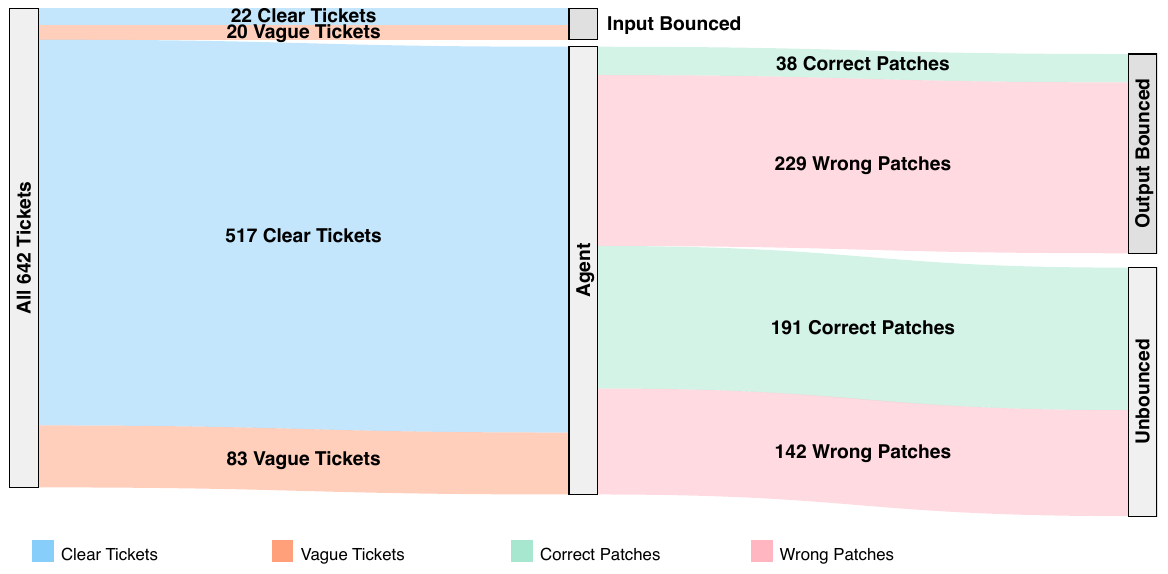}    
    \caption{Sankey diagram depicting effects of the o4-Mini based Input Bouncer and the codex based Output Bouncer on BouncerBench Lite
    }
    \label{fig:sankey}
\end{figure*}

Our experiments show that current LLMs struggle to accurately decide when to reject underspecified inputs~(vague tickets) and incorrect output~(patches). As illustrated by the performance data, even the best-performing reasoning models and agentic systems show only modest effectiveness. This highlights that significant improvements are needed before these systems can be reliably integrated into real-world software engineering workflows. 
While the concept of input and output bouncers is broadly applicable to any autonomous system, the urgency is especially clear in software engineering. The recent surge in LLM-based coding agents \cite{yang2024sweagent, openhands, ruan2024specrover} and their deceptive confidence in suggesting incorrect or low-quality patches makes the need for abstention mechanisms hard to ignore.
This is why we built BouncerBench, to shift focus toward this critical gap and drive progress on systems that can abstain when needed.

We now discuss the effect of using input and output bouncers on BouncerBench Lite. To do this, we consider two scenarios: one without any bouncers, and another where both input and output bouncers are applied.

\subsection{Without Bouncers}

In the first scenario, we assume no filtering is done. The BouncerBench Lite subset contains 642 tickets.
Of these, 103 tickets (16\%) were flagged by annotators as having serious vagueness issues. If we ignore the vagueness issues and pass all 642 tickets to a hypothetical agent that generates the patches we random sampled for BouncerBench, we get 405 incorrect patches. This means 63.1\% of the 642 generated responses are incorrect.

This means the developer would need to manually review and reject or fix 405 of the 642 suggestions. It's worth noting that developers in real-world conditions do not have oracle test cases. The best current LLM-based test generation approach on the SWE-Bench verified split (which only contains well-specified tickets) has an accuracy of 49\% \cite{mundler2024swtbench}, which is much lower than the 70\% ticket resolution rate reported on the same split \cite{jimenez2024swebench}.

\subsection{With Bouncers}

Figure~\ref{fig:sankey} shows the changes when we apply both input and output bouncers.
The input bouncer, using O4-mini, filters out 20 of the 103 underspecified inputs~(vague tickets) before response generation, and incorrectly removing 22 well-specified tickets in the process. Our hypothetical agent now generates 600 patches corresponding to every ticket that isn't input bounced. The output bouncer, using Codex, then further filters the generated responses, rejecting 229 incorrect patches and 38 correct patches. With both bouncers in place, we have 142 wrong patches and 191 correct patches in the final unbounced output. This means 42.6\%~(142 of the 333) responses in the unbounced output are incorrect.

Despite the losses, the reduction in developer review load~(from having to look at 642 responses with 405 incorrect patches to only looking at 333 responses with 142 incorrect patches) is clear. As better bouncers are built, we could reach a state where these 642 tickets are received, but only the 237 correct responses are sent back to the developer.

Table~\ref{tab:bouncer_leaderboard} illustrates what applying our current lightweight bouncers would mean for top submissions on SWE-Bench as compared to the random sample collected for BouncerBench which we discussed above. These correspond to a maximum resolution rate of 33.83\% by SWE-agent followed by Amazon Q at 29.99\% and OpenHands at 29.38\%. We pick submissions from the ``full'' split since we consider all 648 evaluatable cases, 148 of which are not present in the ``verified'' split.
We also note that due to this, the impact of the input bouncer standalone is very muted as most tickets (83.6\%) are well-specified. We however reiterate the need for a separate input bouncer as it would be unreasonable to accurately evaluate the response to a vague question.
Amazon Q and SWE-agent initially provided about half of their patches as correct (50.7\% and 55.9\%, respectively). After deploying both input and output bouncers, these systems showed marked improvements in correctness rates, reaching above 60\%. OpenHands sees similar gains but sticks out as an outlier, losing 30.4\% of its correct patches, however, this high False Positive rate may be attributed to the excessively large patches generated by the tool (10x bigger on average by character count). As we noted while discussing Fig~\ref{fig:ol}, the proposed output bouncers seem to struggle with longer patches. 

\noindent\textbf{Takeaways:} Importantly, abstaining doesn’t mean stopping the pipeline. As shown in Figure~\ref{fig:bouncer_tasks}, underspecified tickets could be routed to refinement modules or flagged for clarification. Rejected incorrect patches could be sent to automated repair steps or dedicated pipelines designed to handle such cases. We see BouncerBench as a starting point for the community to work on this problem and believe building better input and output bouncers should be a key element when designing LLM-based systems. Systems that always act, even when uncertain, can become counterproductive and cause real harm by reinforcing bugs, misleading developers, or wasting review effort.

\begin{table}
\caption{Correct‐patch rates before and after bouncing, average patch length in characters, and valid patches lost on the 648 evaluatable instances}
\centering
\small
\begin{tabular}{lcccc}
\toprule
Submission      & Before (\%) & After (\%) & Chars (avg) & Lost (\%)\\
\midrule
SWE-agent       & 55.9     & 63.8 & 12.0k  & 16.6     \\
Amazon Q        & 50.7     & 61.3 & 1.7k   & 11.3     \\
OpenHands       & 50.5     & 60.7 & 54.1k  & 30.4     \\
\midrule
r\_sample\_all  & 36.9     & 57.4 & 3.8k   & 19.4     \\
\bottomrule
\end{tabular}
\label{tab:bouncer_leaderboard}
\end{table}

\section{Related Work}
\label{sec:related_work}
\noindent\textbf{Software Ticket Resolution Datasets} Over the years several benchmark datasets have been proposed to evaluate repository level bug fixing capabilities. Defects4J \cite{just2014defects4j} focuses on Java and contains only short bug descriptions rather than issue reports. SWE-bench \cite{jimenez2024swebench} was one of the first datasets that mirrored real-world software engineering, in which a code generation system is provided with a codebase and a problem statement, and is tasked to edit that codebase to solve the problem. Several similar datasets have been proposed to address specific concerns, such as the lack of multimodality \cite{yang2024swebenchmultimodal}, specificity to Python \cite{zan2024swe, rashid2025swepolybenchmultilanguagebenchmarkrepository}, understanding economic impact \cite{miserendino2025swe} or security of code \cite{vero2025baxbenchllmsgeneratecorrect}. However, none of these benchmarks have a human verified subset nor were designed to assess a system’s ability to decide when not to process a ticket or withhold an incorrect response, and rather implicitly incentivize always producing something over nothing.
\newline \\
\noindent\textbf{LLM as a judge}
To evaluate and improve agent outputs, researchers are increasingly using LLMs themselves as judges or critics \cite{wang2025-sota-on-swe-bench}. The general LLM-as-a-Judge paradigm treats a large model as an automated evaluator that can flexibly score or critique responses in natural language \cite{zheng2023judging}. Zhao et al. propose CodeJudge-Eval as a benchmark that evaluates the code understanding capabilities of LLMs \cite{zhao2024codejudge}. CodeJudge-Eval consists of competitive programming tasks and thus does not involve real-world tickets or involve full repository context. Building on the idea of LLMs as judges, Zhuge et al. propose an Agent-as-a-Judge framework, an autonomous agent system that evaluates other agents and provides iterative feedback \cite{zhuge2024agent}. The authors also propose a benchmark, DevAI, consisting of 55 AI development tasks with manual annotations that specify a total of 365 hierarchical requirements and find that their framework outperform LLM-as-a-Judge acheiving over 90\% alignment with Human Judges. This benchmark again has no notion of evaluating or bouncing vaguely specified tasks, it is rather aimed at evaluating the ability of judges to check the percentage of explicit binary requirements and preferences met. BouncerBench on the other hand includes real-world tickets from open-source projects that have been annotated by humans to be vague. Further, we do not provide the LLMs with an explicit list of requirements to be assessed, which we consider to be an unrealistic expectation in our setting.
\newline \\
\noindent\textbf{Handling poor report quality}
Real software tickets often suffer from vagueness, missing information, or low quality, which can hinder automated resolution. Prior research has looked at distinguishing bug reports as either good \cite{bettenburg2008makes} or valid \cite{he2020deep} based on patterns in the tickets, and identifying if a bug report lacks key elements like observed vs. expected behavior or steps to reproduce. While the former looks for adequate and correct information, the latter looks at reports that point towards real reproduceable bugs. Fan et al. find the textual features of a bug report and the reporter's experience to be the key distinguishing factors of a valid bug report \cite{fan2018chaff}. Laiq et al. found that these relatively simpler ML techniques worked fairly well in practice with acceptable accuracy but need to be updated at regular intervals to handle concept drift \cite{laiq2024industrial}. With the advent of LLMs, recent work has also looked at enhancing the bug report itself \cite{acharya2025enhancebugreportquality, 10.1145/3691620.3695518}, which represents work that could be leveraged for ``Ticket Refinement'' for cases that are input bounced. While we agree that this line of work on what makes a good bug report can inform better bouncer designs our goal in this study is not to build the best bouncer. Instead, we focus on creating a benchmark that can evaluate abstention and its impact across automated SE workflows, not just bug resolution.
\newline \\
\noindent\textbf{Selecting good responses}
Simply relying on the model’s first output is often insufficient for complex bug fixes. There has been a push towards inference time scaling to achieve SOTA results on SWE-Bench \cite{wang2025-sota-on-swe-bench}. Essentially, we refer to techniques like re-prompting or utilizing multiple agent rollouts followed by a selection mechanism to pick the best solution. Similar to LLM-based APR tools, Agentless \cite{xia2024agentless} proposed sampling multiple potential patches by increasing the temperature while invoking LLMs. It also generated reproduction tests and uses these along with existing regression tests for picking the best patch. OpenHands \cite{openhands} utilizes multiple rollouts to sample patches. In addition to utilizing generated tests like Agentless to filter, it also introduces a critic model that is trained to score the solutions. We note that unlike a bouncer, the critic is utilized here to select the best patch, rather than to abstain. Further, while generating good reproduction tests could help create a better output bouncer, it is not a trivial task. It often requires an executable environment, and the current SOTA LLM-based system dedicated for bug reproduction has an accuracy of just 49\% on the well-specified tickets in SWE-Bench Verified\cite{mundler2024swtbench}.


%

\section{Threats to Validity}
While BouncerBench is built on real-world issues collected by SWE-Bench, it inherits several of its limitations. The benchmark is restricted to Python, focuses on a set of popular GitHub repositories, and only includes issues that are linked to closing PRs which add validation tests. As a result, the performance of bouncers on BouncerBench may not fully generalize to the broader and more diverse range of tickets seen across open-source and industry settings. As such, strong performance here is only an initial signal, not a guarantee of success in more challenging or less curated environments. We also rely on the soundness of the human annotations provided by OpenAI when curating SWE-Bench Verified. In constructing BouncerBench, we take a conservative approach, since it is easy to overlook subtle issues, we ensemble the annotations by selecting the highest-severity label among the three annotators for each instance and treat underspecified inputs~(vague tickets) as unlikely to yield valid solutions.  

Another key issue with utilizing an established benchmark containing historical GitHub issues is data contamination. Recent studies have highlighted concerns about the benchmark being leaked into the training data \cite{zhou2025lessleak, ramos2024large}. We note that every LLM we considered in this study has a knowledge cutoff after the release of SWE-Bench. However, the model with the earliest cutoff of May 2024 (o4-mini) which is before the release of SWE-Bench Verified, performed the best in our evaluations. Further, for input bouncing tasks, as long as the annotations themselves aren't leaked into the training data, it shouldn't help. Output bouncing might benefit if a model has seen the golden patch during training, as this could allow the model to more easily recognize correct or incorrect patches. However, we see no evidence in our results that any of the models are simply memorizing patches while evaluating generated responses that are sourced from a wide range of models. 

We are actively working on integrating newer datasets into BouncerBench that span multiple programming languages and include a larger variety of tasks to address some of these issues and prevent optimizing for a single dataset. While these constraints exist, we believe the benchmark captures key failure modes and provides a practical foundation for more reliable and useful AI assistants in software engineering.

\section{Conclusion}
We introduced BouncerBench, a new benchmark for evaluating whether LLM-based software engineering agents can choose to abstain when inputs are underspecified or outputs are incorrect. Existing systems respond to nearly every input without accounting for uncertainty, which leads to a high rate of incorrect patches and misleading responses. BouncerBench targets this gap by testing agents on their ability to reject underspecified inputs and incorrect outputs.
Our results show that current models, including reasoning models and agents with code access, struggle to bounce cases accurately. Even the best models have limited agreement with human judgments. That said, combining lightweight input and output bouncers does reduce the number of incorrect patches passed to users. These results demonstrate that abstention is a useful but underexplored mechanism for AI-based coding agents and assistants.
By releasing BouncerBench and baseline implementations, we aim to support further research in building more cautious and trustworthy coding agents. The ability to stop and ask for more details when necessary and preventing hallucinated responses from reaching users is essential for reliability, especially as LLMs are deployed more widely in software development.


\balance
\bibliographystyle{IEEEtran}
\bibliography{main}

\begin{thebibliography}{10}
\providecommand{\url}[1]{#1}
\csname url@samestyle\endcsname
\providecommand{\newblock}{\relax}
\providecommand{\bibinfo}[2]{#2}
\providecommand{\BIBentrySTDinterwordspacing}{\spaceskip=0pt\relax}
\providecommand{\BIBentryALTinterwordstretchfactor}{4}
\providecommand{\BIBentryALTinterwordspacing}{\spaceskip=\fontdimen2\font plus
\BIBentryALTinterwordstretchfactor\fontdimen3\font minus \fontdimen4\font\relax}
\providecommand{\BIBforeignlanguage}[2]{{%
\expandafter\ifx\csname l@#1\endcsname\relax
\typeout{** WARNING: IEEEtran.bst: No hyphenation pattern has been}%
\typeout{** loaded for the language `#1'. Using the pattern for}%
\typeout{** the default language instead.}%
\else
\language=\csname l@#1\endcsname
\fi
#2}}
\providecommand{\BIBdecl}{\relax}
\BIBdecl

\bibitem{yang2024sweagent}
\BIBentryALTinterwordspacing
J.~Yang, C.~E. Jimenez, A.~Wettig, K.~Lieret, S.~Yao, K.~R. Narasimhan, and O.~Press, ``{SWE}-agent: Agent-computer interfaces enable automated software engineering,'' in \emph{The Thirty-eighth Annual Conference on Neural Information Processing Systems}, 2024. [Online]. Available: \url{https://arxiv.org/abs/2405.15793}
\BIBentrySTDinterwordspacing

\bibitem{openhands}
\BIBentryALTinterwordspacing
X.~Wang, B.~Li, Y.~Song, F.~F. Xu, X.~Tang, M.~Zhuge, J.~Pan, Y.~Song, B.~Li, J.~Singh, H.~H. Tran, F.~Li, R.~Ma, M.~Zheng, B.~Qian, Y.~Shao, N.~Muennighoff, Y.~Zhang, B.~Hui, J.~Lin, R.~Brennan, H.~Peng, H.~Ji, and G.~Neubig, ``{OpenHands: An Open Platform for AI Software Developers as Generalist Agents},'' 2024. [Online]. Available: \url{https://arxiv.org/abs/2407.16741}
\BIBentrySTDinterwordspacing

\bibitem{ruan2024specrover}
H.~Ruan, Y.~Zhang, and A.~Roychoudhury, ``Specrover: Code intent extraction via llms,'' \emph{arXiv preprint arXiv:2408.02232}, 2024.

\bibitem{rashid2025swepolybenchmultilanguagebenchmarkrepository}
\BIBentryALTinterwordspacing
M.~S. Rashid, C.~Bock, Y.~Zhuang, A.~Buchholz, T.~Esler, S.~Valentin, L.~Franceschi, M.~Wistuba, P.~T. Sivaprasad, W.~J. Kim, A.~Deoras, G.~Zappella, and L.~Callot, ``Swe-polybench: A multi-language benchmark for repository level evaluation of coding agents,'' 2025. [Online]. Available: \url{https://arxiv.org/abs/2504.08703}
\BIBentrySTDinterwordspacing

\bibitem{jimenez2024swebench}
\BIBentryALTinterwordspacing
C.~E. Jimenez, J.~Yang, A.~Wettig, S.~Yao, K.~Pei, O.~Press, and K.~R. Narasimhan, ``{SWE}-bench: Can language models resolve real-world github issues?'' in \emph{The Twelfth International Conference on Learning Representations}, 2024. [Online]. Available: \url{https://openreview.net/forum?id=VTF8yNQM66}
\BIBentrySTDinterwordspacing

\bibitem{mundler2024swtbench}
\BIBentryALTinterwordspacing
N.~M{\"u}ndler, M.~N. Mueller, J.~He, and M.~Vechev, ``{SWT}-bench: Testing and validating real-world bug-fixes with code agents,'' in \emph{The Thirty-eighth Annual Conference on Neural Information Processing Systems}, 2024. [Online]. Available: \url{https://openreview.net/forum?id=9Y8zUO11EQ}
\BIBentrySTDinterwordspacing

\bibitem{mozannar2020consistent}
H.~Mozannar and D.~Sontag, ``Consistent estimators for learning to defer to an expert,'' in \emph{International conference on machine learning}.\hskip 1em plus 0.5em minus 0.4em\relax PMLR, 2020, pp. 7076--7087.

\bibitem{paul2021improving}
R.~Paul, ``Improving the effectiveness of peer code review in identifying security defects,'' in \emph{Proceedings of the 29th ACM Joint Meeting on European Software Engineering Conference and Symposium on the Foundations of Software Engineering}, 2021, pp. 1645--1649.

\bibitem{manning2022research}
S.~Manning, P.~Mishkin, G.~Hadfield, T.~Eloundou, and E.~Eisner, ``A research agenda for assessing the economic impacts of code generation models,'' 2022.

\bibitem{alshahwan2024automated}
N.~Alshahwan, J.~Chheda, A.~Finogenova, B.~Gokkaya, M.~Harman, I.~Harper, A.~Marginean, S.~Sengupta, and E.~Wang, ``Automated unit test improvement using large language models at meta,'' in \emph{Companion Proceedings of the 32nd ACM International Conference on the Foundations of Software Engineering}, 2024, pp. 185--196.

\bibitem{weisz2025examining}
J.~D. Weisz, S.~V. Kumar, M.~Muller, K.-E. Browne, A.~Goldberg, K.~E. Heintze, and S.~Bajpai, ``Examining the use and impact of an ai code assistant on developer productivity and experience in the enterprise,'' in \emph{Proceedings of the Extended Abstracts of the CHI Conference on Human Factors in Computing Systems}, 2025, pp. 1--13.

\bibitem{liu2024exploring}
F.~Liu, Y.~Liu, L.~Shi, H.~Huang, R.~Wang, Z.~Yang, L.~Zhang, Z.~Li, and Y.~Ma, ``Exploring and evaluating hallucinations in llm-powered code generation,'' \emph{arXiv preprint arXiv:2404.00971}, 2024.

\bibitem{sergeyuk2025using}
A.~Sergeyuk, Y.~Golubev, T.~Bryksin, and I.~Ahmed, ``Using ai-based coding assistants in practice: State of affairs, perceptions, and ways forward,'' \emph{Information and Software Technology}, vol. 178, p. 107610, 2025.

\bibitem{murphy2011we}
E.~Murphy-Hill, C.~Parnin, and A.~P. Black, ``How we refactor, and how we know it,'' \emph{IEEE Transactions on Software Engineering}, vol.~38, no.~1, pp. 5--18, 2011.

\bibitem{johnson2023make}
B.~Johnson, C.~Bird, D.~Ford, N.~Forsgren, and T.~Zimmermann, ``Make your tools sparkle with trust: The picse framework for trust in software tools,'' in \emph{2023 IEEE/ACM 45th International Conference on Software Engineering: Software Engineering in Practice (ICSE-SEIP)}.\hskip 1em plus 0.5em minus 0.4em\relax IEEE, 2023, pp. 409--419.

\bibitem{kamath2020selective}
A.~Kamath, R.~Jia, and P.~Liang, ``Selective question answering under domain shift,'' \emph{arXiv preprint arXiv:2006.09462}, 2020.

\bibitem{jurayj2025your}
W.~Jurayj, J.~Cheng, and B.~Van~Durme, ``Is that your final answer? test-time scaling improves selective question answering,'' \emph{arXiv preprint arXiv:2502.13962}, 2025.

\bibitem{verma2023learning}
R.~Verma, D.~Barrej{\'o}n, and E.~Nalisnick, ``Learning to defer to multiple experts: Consistent surrogate losses, confidence calibration, and conformal ensembles,'' in \emph{International Conference on Artificial Intelligence and Statistics}.\hskip 1em plus 0.5em minus 0.4em\relax PMLR, 2023, pp. 11\,415--11\,434.

\bibitem{openaiIntroducingSWEbench}
``{I}ntroducing {S}{W}{E}-bench {V}erified --- openai.com,'' \url{https://openai.com/index/introducing-swe-bench-verified}, [Accessed 30-05-2025].

\bibitem{annotationInstructions}
\url{https://cdn.openai.com/introducing-swe-bench-verified/swe-b-annotation-instructions.pdf}, [Accessed 30-05-2025].

\bibitem{OpenAI}
\url{https://openai.com/index/o3-o4-mini-system-card}, [Accessed 30-05-2025].

\bibitem{Claude}
\url{https://www.anthropic.com/news/claude-3-7-sonnet}, [Accessed 30-05-2025].

\bibitem{yang2025qwen3}
A.~Yang, A.~Li, B.~Yang, B.~Zhang, B.~Hui, B.~Zheng, B.~Yu, C.~Gao, C.~Huang, C.~Lv \emph{et~al.}, ``Qwen3 technical report,'' \emph{arXiv preprint arXiv:2505.09388}, 2025.

\bibitem{team2025gemma}
G.~Team, A.~Kamath, J.~Ferret, S.~Pathak, N.~Vieillard, R.~Merhej, S.~Perrin, T.~Matejovicova, A.~Ram{\'e}, M.~Rivi{\`e}re \emph{et~al.}, ``Gemma 3 technical report,'' \emph{arXiv preprint arXiv:2503.19786}, 2025.

\bibitem{gu2024survey}
J.~Gu, X.~Jiang, Z.~Shi, H.~Tan, X.~Zhai, C.~Xu, W.~Li, Y.~Shen, S.~Ma, H.~Liu \emph{et~al.}, ``A survey on llm-as-a-judge,'' \emph{arXiv preprint arXiv:2411.15594}, 2024.

\bibitem{just2014defects4j}
R.~Just, D.~Jalali, and M.~D. Ernst, ``Defects4j: A database of existing faults to enable controlled testing studies for java programs,'' in \emph{Proceedings of the 2014 international symposium on software testing and analysis}, 2014, pp. 437--440.

\bibitem{yang2024swebenchmultimodal}
\BIBentryALTinterwordspacing
J.~Yang, C.~E. Jimenez, A.~L. Zhang, K.~Lieret, J.~Yang, X.~Wu, O.~Press, N.~Muennighoff, G.~Synnaeve, K.~R. Narasimhan, D.~Yang, S.~I. Wang, and O.~Press, ``{SWE}-bench multimodal: Do ai systems generalize to visual software domains?'' in \emph{The Thirteenth International Conference on Learning Representations}, 2025. [Online]. Available: \url{https://openreview.net/forum?id=riTiq3i21b}
\BIBentrySTDinterwordspacing

\bibitem{zan2024swe}
D.~Zan, Z.~Huang, A.~Yu, S.~Lin, Y.~Shi, W.~Liu, D.~Chen, Z.~Qi, H.~Yu, L.~Yu \emph{et~al.}, ``Swe-bench-java: A github issue resolving benchmark for java,'' \emph{arXiv preprint arXiv:2408.14354}, 2024.

\bibitem{miserendino2025swe}
S.~Miserendino, M.~Wang, T.~Patwardhan, and J.~Heidecke, ``Swe-lancer: Can frontier llms earn 1 million from real-world freelance software engineering?'' \emph{arXiv preprint arXiv:2502.12115}, 2025.

\bibitem{vero2025baxbenchllmsgeneratecorrect}
M.~Vero, N.~Mündler, V.~Chibotaru, V.~Raychev, M.~Baader, N.~Jovanović, J.~He, and M.~Vechev, ``Baxbench: Can llms generate correct and secure backends?'' 2025.

\bibitem{wang2025-sota-on-swe-bench}
\BIBentryALTinterwordspacing
X.~Wang, ``Sota on swe-bench verified with inference-time scaling and critic model,'' \emph{All Hands AI Blog}, April 2025. [Online]. Available: \url{https://www.all-hands.dev/blog/sota-on-swe-bench-verified-with-inference-time-scaling-and-critic-model}
\BIBentrySTDinterwordspacing

\bibitem{zheng2023judging}
L.~Zheng, W.-L. Chiang, Y.~Sheng, S.~Zhuang, Z.~Wu, Y.~Zhuang, Z.~Lin, Z.~Li, D.~Li, E.~Xing \emph{et~al.}, ``Judging llm-as-a-judge with mt-bench and chatbot arena,'' \emph{Advances in Neural Information Processing Systems}, vol.~36, pp. 46\,595--46\,623, 2023.

\bibitem{zhao2024codejudge}
Y.~Zhao, Z.~Luo, Y.~Tian, H.~Lin, W.~Yan, A.~Li, and J.~Ma, ``Codejudge-eval: Can large language models be good judges in code understanding?'' \emph{arXiv preprint arXiv:2408.10718}, 2024.

\bibitem{zhuge2024agent}
M.~Zhuge, C.~Zhao, D.~Ashley, W.~Wang, D.~Khizbullin, Y.~Xiong, Z.~Liu, E.~Chang, R.~Krishnamoorthi, Y.~Tian \emph{et~al.}, ``Agent-as-a-judge: Evaluate agents with agents,'' \emph{arXiv preprint arXiv:2410.10934}, 2024.

\bibitem{bettenburg2008makes}
N.~Bettenburg, S.~Just, A.~Schr{\"o}ter, C.~Weiss, R.~Premraj, and T.~Zimmermann, ``What makes a good bug report?'' in \emph{Proceedings of the 16th ACM SIGSOFT International Symposium on Foundations of software engineering}, 2008, pp. 308--318.

\bibitem{he2020deep}
J.~He, L.~Xu, Y.~Fan, Z.~Xu, M.~Yan, and Y.~Lei, ``Deep learning based valid bug reports determination and explanation,'' in \emph{2020 IEEE 31st International Symposium on Software Reliability Engineering (ISSRE)}.\hskip 1em plus 0.5em minus 0.4em\relax IEEE, 2020, pp. 184--194.

\bibitem{fan2018chaff}
Y.~Fan, X.~Xia, D.~Lo, and A.~E. Hassan, ``Chaff from the wheat: Characterizing and determining valid bug reports,'' \emph{IEEE transactions on software engineering}, vol.~46, no.~5, pp. 495--525, 2018.

\bibitem{laiq2024industrial}
M.~Laiq, N.~b. Ali, J.~B{\"o}rstler, and E.~Engstr{\"o}m, ``Industrial adoption of machine learning techniques for early identification of invalid bug reports,'' \emph{Empirical Software Engineering}, vol.~29, no.~5, p. 130, 2024.

\bibitem{acharya2025enhancebugreportquality}
\BIBentryALTinterwordspacing
J.~Acharya and G.~Ginde, ``Can we enhance bug report quality using llms?: An empirical study of llm-based bug report generation,'' 2025. [Online]. Available: \url{https://arxiv.org/abs/2504.18804}
\BIBentrySTDinterwordspacing

\bibitem{10.1145/3691620.3695518}
\BIBentryALTinterwordspacing
L.~Bo, W.~Ji, X.~Sun, T.~Zhang, X.~Wu, and Y.~Wei, ``Chatbr: Automated assessment and improvement of bug report quality using chatgpt,'' in \emph{Proceedings of the 39th IEEE/ACM International Conference on Automated Software Engineering}, ser. ASE '24.\hskip 1em plus 0.5em minus 0.4em\relax New York, NY, USA: Association for Computing Machinery, 2024, p. 1472–1483. [Online]. Available: \url{https://doi-org.proxy.lib.uwaterloo.ca/10.1145/3691620.3695518}
\BIBentrySTDinterwordspacing

\bibitem{xia2024agentless}
C.~S. Xia, Y.~Deng, S.~Dunn, and L.~Zhang, ``Agentless: Demystifying llm-based software engineering agents,'' \emph{arXiv preprint arXiv:2407.01489}, 2024.

\bibitem{zhou2025lessleak}
X.~Zhou, M.~Weyssow, R.~Widyasari, T.~Zhang, J.~He, Y.~Lyu, J.~Chang, B.~Zhang, D.~Huang, and D.~Lo, ``Lessleak-bench: A first investigation of data leakage in llms across 83 software engineering benchmarks,'' \emph{arXiv preprint arXiv:2502.06215}, 2025.

\bibitem{ramos2024large}
D.~Ramos, C.~Mamede, K.~Jain, P.~Canelas, C.~Gamboa, and C.~L. Goues, ``Are large language models memorizing bug benchmarks?'' \emph{arXiv preprint arXiv:2411.13323}, 2024.

\end{thebibliography}
\end{document}